\documentclass[sn-basic]{sn-jnl}

\usepackage{graphicx}\usepackage{xcolor}
\usepackage{multirow}\usepackage{amsmath,amssymb,amsfonts}\usepackage{amsthm}\usepackage{mathrsfs}\usepackage[title]{appendix}\usepackage{xcolor}\usepackage{textcomp}\usepackage[capitalise]{cleveref}
\usepackage{booktabs}
\usepackage{makecell}
\usepackage{adjustbox}
\usepackage{tikz}
\usepackage{comment}

\usepackage[
 disable,
 textsize=footnotesize,linecolor=teal,bordercolor=teal,backgroundcolor=yellow!10]{todonotes}
\newcommand{\quotes}[1]{``#1''}
\newcommand{\prt}[2]{\prto{#1.}{#2.}}
\newcommand{\prto}[2]{\textit{#1#2}}
\newcommand{\framework}{PIRA}
\makeatletter
\if@todonotes@disabled
\else \paperwidth=\dimexpr \paperwidth + 6cm\relax
\oddsidemargin=\dimexpr\oddsidemargin + 3cm\relax
\evensidemargin=\dimexpr\evensidemargin + 3cm\relax
\marginparwidth=\dimexpr \marginparwidth + 3cm\relax

\fi
\makeatother

\begin{document}

\phantom{a}
\vfil
{\Large
\begin{center}
\textbf{A Mixed-Method Framework for Evaluating the Social Impact of Community Cooperation Projects in Developing Countries}
\end{center}}
\vspace{6em}

\begin{center}

Giorgia Sampò\\
University of Southern Denmark, Denmark\\
gisa@sam.sdu.dk
\\[2em]

Saverio Giallorenzo\\
Alma Mater Studiorum - Universit\`a di Bologna, Italy\\
INRIA, France\\
saverio.giallorenzo@unibo.it
\\[2em]

Zelda Alice Franceschi\\
Alma Mater Studiorum - Universit\`a di Bologna, Italy\\
zelda.franceschi@unibo.it
\end{center}

\title{}

\abstract{Why do some community-cooperation projects catalyse participation 
through durable, resilient collaboration networks while others result in 
negligible impact and leave the local social fabric unchanged? We argue outcomes
hinge on participation architecture: simple, visible routines -- onboarding
help, templated tasks, lightweight contribution/benefit tracking -- that create
easy ``entry portals'' and route work across clusters without heavy hierarchy.
We introduce Project Intervention Response Analysis (PIRA), a mixed
anthropological-network-analysis framework that compares observed community
networks with counterfactual networks absent from project-induced ties. PIRA
also adds a new egocentric metric to detect ``architectural alters'' -- latent
facilitators and boundary spanners. We begin validating PIRA in a three-month
field study in Pomerini, Tanzania, where NGOs coordinated citizens,
associations, and specialists.
Findings indicate that sociotechnical participation architectures -- not
charismatic hubs -- underwrite durable coordination. PIRA offers a reusable
method to link organizational design mechanisms to formal network signatures.}


\maketitle

\section{Introduction}
\label{sec:intro}

Communities in developing countries often face significant challenges, such as
poverty and economic instability~\citep{K86}, social fragmentation~\citep{O11},
and weak institutional structures~\citep{TP11,RA12}, that can deeply affect the
resilience of their social fabric as well as the impact of institutional
organisations. In these contexts, non-governmental organisations (NGOs) foster a
counter-balancing effect, implementing projects aimed at enforcing community
cooperation and rebuilding social connections~\citep{MR12}. However, these
interventions occur in complex social environments where the community and
organisational structures are simultaneously evolving, influencing each other's
development and direction~\citep{LN83,GV91,MB9,MLG11}.

The effectiveness of international cooperation, and specifically community
cooperation (i.e., a setup characterised by small scale interventions and high
involvement of community members throughout planning and development), heavily
relies on the organisations' ability to functionally interact with the existing
cultural and historical environment they are trying to act within~\citep{LN83}.
The fine-tuning of organisational practices on the local setting of action
becomes ever more relevant in the case of community
cooperation~\citep{Colajanni_1994}, where the main goal of projects is to
reinforce and leverage the social fabric, fostering local involvement  
in and through ``self-centred development'', where interventions are ideated,  
supported, and steered by the community itself. Community cooperation projects
therefore become  ``collections of actors whose membership in the collective
provides social and cultural resources that shape their action''~\citep{MLG11}.
The entangling of organisations in the fabric of the community benefitting from
interventions affects the set-up and direction of projects as well as the
conceptual and practical aspects of their management.

Despite studies recognising the reciprocal influence of the targeted
socio-cultural environment and the organisational structure(s) supporting
interventions~\citep{AMC17}, there are no systematic frameworks for
understanding if and how the emergence and adaptation of community cooperation
organisations can reinforce community relations, and how these organisations
are, in their turn, moulded and bent by those same communities. Moreover,
research considering both social and organisational factors~\citep{AMC17, MF07}
provide limited support for addressing the open question of how organisations
(and, in our case, interventions) and community structures co-evolve,
particularly in contexts where both the social fabric and organisational
structures are fluid and emerging. This scholarly gap is especially evident in
situations where community relations are not yet solidified or are undergoing
significant transformation. 

Furthermore, while existing literature primarily focuses either on how
communities influence organisational initiatives~\citep{MLG11,OL11,GA23, MB9,
LN83} or on how organisations reshape communities~\citep{B12,MW03,RGAP16}, there
are no works that address these aspects in unison. Similarly, while the
literature explores how organisations implement development projects, these
proposals emphasise how interventions address the needs of established
communities~\citep{BF06,ASH07,NR10,I16} rather than how these projects
simultaneously build community relations while evolving their own organisational
structures. Therefore, explaining the different processes by which community
relations and intervention structures co-emerge is both theoretically and
practically important.

Thus, we ask: 

\vspace{1em}
\begin{center}
\emph{How do community cooperation interventions influence the relations among
projects' participants and their acquaintances while simultaneously shaping the
organisational structures implementing these interventions?} 
\end{center}
\vspace{1em}

To address our research question, we propose a framework called ``Project
Interventions Response Analysis'' (\framework), to measure the effects of
community cooperation interventions on the social fabric of target communities,
and vice versa, enabling longitudinal evaluations of projects design and
effectiveness. We base PIRA on a mixed-method approach, combining social network 
analysis (SNA) with ethnographic research. This perspective is informed by our
analysis of published case studies, where we found common denominators of latent
general methods for studying communities in developing societies and how NGO
interventions impact their social fabric, along with a general lack of
systematic and comprehensive solutions and tools of analysis.

Inspired by these findings, we define our framework to include established
measures from the literature -- reviewed in \cref{sec:state_of_the_art} -- and
introduce a \emph{novel measure for analysing the importance of peripheral
community members in determining network connectedness} -- in the context of
ego-network measures, where the network is divided between the egos, which are
the study's participants, and the alters, which are non-participants mentioned
by the egos. We explain the components and phases that characterise \framework{}
in \cref{sec:framework}, rooting their origins in anthropological studies
(interested in qualitative interviews and questionnaires viewed through a
contextualised interpretation) and SNA (focused on quantitative measures that
help unearth information at both personal and community levels).

In \cref{sec:case_study}, we validate \framework{} through a case study in
Pomerini, a village in the Tanzanian highlands, conducted over a period of 3
months and examining 382 respondents' relations before and after project
participation. From the analysis, which we discuss in \cref{sec:discussion}, we
observe a positive effect of interventions in consolidating the social fabric of
the village. However, we also note that NGOs could better coordinate and involve
local figures who emerged as central both quantitatively- (they bridge different
groups and/or have many connections) and personality-wise (for the role they
play in the community). More in general, we report three main meta-findings
related to our proposal. First, our framework explains how interventions and
communities co-establish meaning and collective goals. Second, it can reveal
patterns of how interventions reshape community relations while simultaneously
structuring their own organisational processes. Third, it can indicate the
degree to which interventions embed themselves in the communities they are
(re)building, explaining their varying influence on community relations and
access to resources.

We make publicly available at \url{https://zenodo.org/records/11544015} all the
data and source code of the case study: the structure of the interviews (both in
English and Swahili) and their anonymised transcript, the questionnaire (both in
English and Swahili) and its anonymised results, the network datasets, and the
implementation of the novel measure introduced with \framework{} to facilitate
further research and practical application. 
\section{State of the Art}
\label{sec:state_of_the_art}

This work is not the first one that sees a mixture of Anthropology and
SNA~\citep{DH14,CSW05}, e.g., researchers integrated SNA with ethnographic
research for data gathering~\citep{RuralChina,FreemanVisualization} or
complemented anthropological analysis with SNA
investigation~\citep{anthropologyDataMining,Knox_Savage_harvey}. However, to the
best of our knowledge, no proposal mixes anthropology and SNA for studying
community cooperation. Thus, in the following, we focus on related work around
community cooperation from either perspective and conclude by looking at the 
emerged aspects from the point of view of organisational theory.

\subsection{Community Cooperation and Analysis of Anthropological Settings}

The branch of international cooperation dubbed \emph{community cooperation}
advocates the direct involvement of communities in project ideation and
realisation. Hence, community cooperation focuses on self-development and local
involvement, integrating policies and technologies based on local culture and
traditions. This approach leads to smaller-scale interventions like rural or
village projects, departing from the ineffective imposition of ``good
practices'' on perceived underdeveloped countries~\citep{Colajanni_1994}.

The anthropological approach, encompassing both a global view and the specific
behaviours of individuals in mental, conceptual, and symbolic aspects, allows
the researcher to address the cultural dimensions of development -- in
particular, connecting the anthropological concept of culture as the entire
patrimony of techniques, institutions, costumes, ideas, beliefs, and values of a
society.

According to Brewer, data collection methods shall capture the ``social meanings
and ordinary activities'' of people (informants) in ``naturally occurring
settings'' -- i.e., ``the field''~\citep{Brewer}. Studies can consider secondary
research and document analysis for insights into the research topic, e.g.,
kinship charts, used in anthropological studies to help uncover logical patterns
and social structures in non-Western societies.

Depending on the aim of the study, the anthropological perspective on community
cooperation either corresponds to Colajanni's ``development processes
anthropology'' or ``anthropology for development
processes''~\citep{Colajanni_1994}. The distinction respectively lies in the
willingness of the anthropologist to investigate the situation in isolation or
direct interventions and personnel formation on the field. Hence, in the first
case, the study aims to further research, while in the second the results inform
and direct personnel.

Methodologically, the collection and interpretation of data by an anthropologist
inevitably introduce bias in studies, threatening their reliability.
Ethnographers frequently mitigate this issue using a principle called
``reflexivity'', where they strive to ``explore [how the] researcher's
involvement with a particular study influences, acts upon and informs such
research''~\citep{Nightingale_Cromby}. Nonetheless, the researcher's perspective
is a critical aspect of the discipline, influencing sampling processes and the
choice of key informants. Indeed, due to time/resource constraints, researchers
often consider a subset of the individuals in the community under study, making
it crucial to scrutinise the selection process and the potential representation
biases therein.

We deem the above methods important to provide contextual knowledge for the
researcher in qualitatively evaluating the evidence from quantitative analyses
and to identify participants for the quantitative data gathering. Specifically,
\framework{} includes participant observation, field notes, interviews, and
surveys.

\subsection{Social Network Analysis and Social Capital}

From the SNA perspective, analysing the relationships, trust, and cooperation
networks within a community condensates in the study of \emph{social capital}.
In the literature, we find two main definitions of social capital. Focussing on
groups, social capital is ``partly cultural, partly socio-structural including
such things as [the] rule of law, social integration, and
trust''~\citep{Borgatti_Connections}. Looking at individuals, we have ``the
value of an individual's social relationships''~\citep{Burt}, possibly seen as a
source of power. To build \framework{}, we follow the second definition and draw
part of \framework{}'s network measures from related published studies,
discussed below.

Marques and Bichir study the ``Poverty and Sociability in Brazilian
Metropo\-lises''~\citep{Bichir_Marques}. The authors start with the assumption
-- based on preliminary studies -- that the role of personal networks influences
the diffusion of urban poverty in metropolitan cities, identifying sociability
as a central issue when analysing urban poverty. What they found was that ``more
local and more \emph{homophilic} networks are associated with worse social
conditions'', aligning with the definition presented by Borgatti et al.\@ that
homophily can have negative connotations, e.g., providing less exposure to a
wide range of ideas~\citep{Borgatti_Connections}.

García-Amado et al.\@ studied the social capital of a forest community in a
biosphere reserve in Chiapas, Mexico~\citep{GarciaAmado}, adopting the
definition of social capital as ``means for fostering collective action by
lowering transaction costs and inhibiting
free-riding''~\citep{Lehtonen,Ostrom,Putnam}. The authors analysed the community
of ``ejidatarios'' -- landowners with full voting and decision rights in
assemblies -- and ``pobladores'' -- workers without land rights, except for
small portions donated/sold by ejidatarios. From the study, ejidatarios share a
high level of \emph{centrality} and \emph{in-degree}, owning resources and
knowledge. Moreover, new activities (e.g., opening a coffee market where
publadores can participate) have low network \emph{transitivity}, while older,
more regulated ones, like palm agriculture, have higher values due to higher
planning and regulation requirements, i.e., transitivity detects the different
networks as lax and inclusive vs tight and selective.
 
Mertens et al.\@ analyse ``[t]he role of strong-tie social networks in mediating
food security of fish resources by a traditional riverine community in the
Brazilian Amazon''~\citep{Mertens}. The authors structured their analysis
through a network approach looking at availability, access, usage, and stability
of food. Following Wellman and Wortley's interpretation that ``along with market
exchanges and the institutional distribution of services and goods, social ties
and networks are a significant way through which people obtain
resources''~\citep{Wellman_Wortley}, the authors study the
\emph{relationships} among villagers and the related \emph{clustering} of the
network.

Pietri et al.\@ investigate the effects of the Coral Triangle~\citep{Pietri} --
a regional learning network originated from a marine governance initiative to
foster regional exchanges -- in facilitating ``capacity building'' and enable
individuals to ``strengthen skills, knowledge, and relationships to promote the
realisation of joint goals''.
To study the network, the researchers measure \emph{density}, \emph{average
degree}, \emph{degree centralisation}, \emph{fragmentation} (both directed and
reciprocal ties), and the \emph{brokerage score} of the individuals, to
precisely define their network roles. They found that \quotes{[i]f learning
networks have an opportunity for a local coordinating body, such an entity may
be better positioned to emphasise the importance of local knowledge, encourage
social learning, and sustain network functions. Learning networks should also
stress to members the importance of broadening the network and disseminating
lessons learned to peers in their professional communities who are not learning
network participants. [...] The tangible learning and capacity development
outcomes cultivated through the [...] network underscore the value of and need
to invest in conservation networks that support peer-to-peer social
learning}~\citep{Pietri}.

\subsection{Organisational Settings}

Going beyond classic community-driven development, participation architecture
research highlights how simple, visible rules and routines (onboarding scripts,
templated tasks, transparent benefit allocation) can stabilise self-management
while keeping decision rights local. Massa and O'Mahony~\citep{MM21} show how
such sociotechnical scaffolding supplants normative control in open communities;
Gil et al.~\citep{MS23} extend the insight to hybrid settings in which
hierarchies catalyse and support self-managed groups via points-based systems.
This fact complements an anthropological stance: interventions are not only
``projects'' but normative infrastructures that communities appropriate, adapt,
and embed.

There has been significant research on how NGOs influence communities and how
existing community structures shape organisational
interventions~\citep{W90,MO02,BLRS18}. In particular, we look at research on
self-managed collectives~\citep{MS23} showing that hierarchical social movement
organisations can seed and harness these structures through point-based
participation architectures that align local benefits with higher-order policy
goals. Reading NGO interventions through this lens, we treat them as
opportunities to design or strengthen in-situ decentralised participation
architectures. 

At the same time, work on participation architectures (i.e., the designed
configurations of people, technology, and rules that enable and structure
participation in collective activities) shows how communities pursuing
collective ends can scale collaboration without reverting to top-down
control~\citep{MM21} -- drawing a parallel with how digital activists can create
a decentralised sociotechnical scaffolding through onboarding spaces, templates,
and routing mechanisms that channel newcomers' contributions and coordinate
interdependent tasks at scale.

Drawing these results back to an organisational perspective, it emerges that
participation architectures suggest testable network footprints. By multiplying
local ``portals'' for participation and standardising hands-off, successful
projects should strengthen the social fabric under intervention by lowering (i)
its betweenness centralisation (many micro-brokers instead of a few) and (ii)
core-periphery contrast, while increasing (iii) its heterophilic bridging ties
across groups and (iv) the salience of architectural alters -- individuals or
roles (often facilitators) that sit outside the formal participant set yet
mediate ego-ego connectivity.

Building on this body of evidence, \framework{}establishes an integrated
framework for analysing and quantifying the social, organisational, and network
dimensions of collaboration architectures. The framework enables in-situ
examination of structural properties such as centralisation, bridging ties, and
the function of architectural ``gateways'' and intra-organisational boundary
spanners. While \framework{}draws on established anthropological approaches and
network metrics to address these dimensions, it also advances the literature by
introducing a novel measure -- \emph{2-Ego Link Dealt by Alter} (2-ELDA)
(cf.~\cref{sec:ZELDA}) -- which supports the systematic identification and
quantification of salient architectural alters.

Conceptually, we argue that 2-ELDA operationalises what organisational theorists
have long described as \emph{boundary spanning}~\citep{TU81,LE5}, i.e., the
activity of linking otherwise segregated subunits, professional groups, or
knowledge domains. Traditional network measures such as betweenness or
structural holes~\citep{Burt} provide partial insights into such brokerage, but
they are typically egocentric or path-based. 2-ELDA complements this tradition
by focusing on alters' role as a second-order broker -- e.g., as individuals who
connect egos who are themselves not directly linked through peers. This
perspective captures a distinctive form of architectural
brokerage~\citep{OB5,CU5} that is theoretically salient but, to the best of our
knowledge, has been difficult to conceptually define as an objective measure.

For organisational scholars, 2-ELDA enables the systematic identification of
individuals or units who sustain cross-boundary connectivity within
collaboration networks -- be it between departments, disciplines, or
organisations. Such actors are crucial for organisational learning and
adaptation~\citep{CR4,CA2}, as they channel information across structural
divides while maintaining the integrity of local communities. By quantifying
this bridging role, 2-ELDA allows researchers to move beyond qualitative
recognition of boundary spanners toward objective, comparative, dynamic, and
large-scale analyses of how collaborative architectures evolve and where
coordination bottlenecks or innovation opportunities arise. 
\section{A Framework for Analysing Community Cooperation Interventions}
\label{sec:framework}

We now introduce \framework{}. The practical goal of \framework{} is to assess
the influence on the community fabric of people's participation in cooperation
interventions. In \framework{}, we draw this assessment by comparing the network
of the community with and without the relationships induced by participation in
interventions. Moreover, during PIRA's qualitative phase, we explicitly track
organisational architectural features while, during the quantitative phase, we
examine whether these features coincide with structural-level changes
anticipated by the participation-architecture lens (cf.~\cref{sec:ZELDA}), using
2-ELDA to surface architectural alters who mediate ego-ego ties despite not
being formal respondents.

In PIRA, we use anthropological analysis for two main purposes. First, it
provides the researcher with preliminary knowledge about the social values and
dynamics of the community under study, which informs the different aspects of
the collection of quantitative data (framing of relationships within the
cultural settings of the community, identification of the study's participants,
etc.). Second, it supports the interpretation of quantitative data and its
results with insights gathered from the observed social dynamics. We use SNA to
represent the social fabric of the community under study as nodes -- the people
in the community -- and edges -- the relationships among the people in the
community -- and apply measures to extract quantitative information on their
configurations (e.g., central figures and sub-communities).

Specifically, \framework{} provides an overview of the social dynamics of the
community subject of projects, in particular, identifying key figures to improve
the ideation and deployment of projects. Moreover, \framework{} works as a
platform to monitor an ongoing (set of) project(s) w.r.t.\@ its social outcomes
(e.g., increase social cohesion). A researcher can use \framework{} to assess
the social effects of a given set of interventions on a community (within a set
timeframe) by applying the framework's measures and interpreting their results
from the contextual reading given below.

\framework{} encompasses 6 steps, divided into 3 phases, detailed below. In
summary, the first phase, of data collection (steps 1 and 2), applies principles
and techniques from anthropology to acquire a general, qualitative knowledge of
the community and population under study, identifying the candidates who could
take part in the study, and contextualising the following research steps within
the societal and cultural settings of the studied population. The second phase
(steps 3 and 4) regards data digitisation and cleaning. The third phase (steps 5
and 6) concerns the quantitive analysis of the data and its qualitative reading,
complemented by the first-phase knowledge.

\begin{enumerate}

    \item \emph{Participant observation}: gather qualitative information
    regarding the study setting. We connotate this step (like the next one) with
    ethnographic or sociological methodologies and tools. We suggest starting
    the observation with meaningful figures (e.g., people in charge of projects,
    participants with specific traits/positions, etc.) to guide the researcher
    in the discovery of the local culture. As per diligence (cf.
    \cref{sec:state_of_the_art}), the researcher should use the information that
    emerged from the observation of the population to identify a set of key
    informants useful for later investigations -- e.g., interviews about social
    dynamics that appear as obscure and that might assume importance for the
    phenomenon under analysis. Due diligence also includes keeping a field
    journal during the whole process, to later recollect thoughts and
    observations which can help in contextualising and interpreting the results.

    \item \emph{Semi-structured interviews}: interview the key informants
    identified in the previous step regarding both meaningful dynamics that are
    impossible to observe -- due to time or social limitations -- and
    experiences or perceptions acquirable only through the recollections of
    participants. An example encompassing both cases could be asking the key
    informants to describe how they first met their friends, and what they
    usually do together. The rationale behind this question is both to
    understand the dynamics of social aggregation and to gather a shared
    definition of the relationships under study. Examples of possible subjects
    of semi-structured interviews are family relationships (e.g., what do they
    define a normal family structure? Who do they consider a family member? Who
    is a relative?), social dynamics (e.g., what are the main occasions to meet
    people? What traits do they attribute to a friendship relation?), and time
    scheduling and daily activities (e.g., what do people do during their day?
    Which part of the day do they devote to work? What do people do in their
    spare time?).
    
    Besides having to confront with a different cultural frame, the researcher
    might need to conduct the interviews in an unfamiliar language. Thus, we
    recommend working with cultural mediators who are native speakers of the
    language used in the interviews and familiar with the language the
    researcher uses to define the interview questions, to ensure a high
    correspondence between the intentions of the questions and the translation
    (both questions and answers).

    This step aims to provide the researcher with integral and accurate insights
    into the culture underlying the social structure. The step is fundamental
    also for defining a meaningful set of questions for the next step of the
    administration of questionnaires for the collection of quantitative data.
    
    \item \emph{Administration of questionnaires}: collect quantitative data on
    the target social relations. The development of this phase follows from the
    results of the previous one, in particular, regarding the development of a
    questionnaire used to elicit quantitative data from the respondents.
    
    After having built the questionnaire in the language of the researcher's
    choice, they might need to prepare a translated version (more than one,
    depending on the number of native languages spoken by the studied
    population). The translation process must ensure the highest possible
    correspondence between the version prepared by the researcher and its
    localisations. An important point for the translation is that the
    \emph{meaning} of the questions is idiomatically adapted and culturally
    contextualised in the target language rather than having a word-by-word
    translation.
    
    From the practical point of view, the researcher shall also perform the
    administration of questionnaires in person, whenever possible, to allow them
    to annotate meaningful behaviour (e.g., discomfort towards or difficulties
    in understanding particular questions) or contextual information (e.g., the
    spatial proximity of respondents, which might help in clarifying information
    such as monikers and abbreviations) and take notes of the settings (e.g.,
    mood, proximity of respondents).

    While \framework{} does not fix a specific structure for the questionnaires,
    from our experience (and the conduction of the case study,
    cf.~\cref{sec:case_study}) we suggest using a mix of open and closed
    questions (e.g., open list of names, closed multiple-choice questions). In
    particular, we suggest the use of open questions to let the participants
    freely mention people they think belong to non-predetermined groups (e.g.,
    family, friends, co-workers). Closed questions increase recall and
    decrease omissions, e.g., when stating the participation in projects and to
    collect perceptual data thereof.

    An invariant is including a question on what relationships a
    respondent has developed from the participation in interventions (project
    activities, planning, etc.).

    \item \emph{Data cleaning (and digitisation)}: if, in the previous phase,
    the collected data is in a non-digital format (e.g., questionnaire printed
    on paper), the researcher shall digitise it. The pros of digitising the
    gathered data are multiple. The main ones include simplifying storage and
    duplication (backups) and the possibility to both define standardised
    routines to clean the data and automatise the application of quantitative
    measures. Suitable formats for the digitisation of the questionnaires
    towards the next steps are comma-separated values (CSV) and other tabular
    formats.

    Given the digital data, the researcher applies a check-and-cleaning routine
    useful for consolidating it. Possible verifications and interventions
    include the correction of typos and the reconciliation of inconsistencies.
    For reliability, the researcher shall report on the kind of procedures they
    used to clean the data. These include trivial activities, like spelling
    correction, but also more involved ones, e.g., the definition and
    application of heuristics that deal with the disambiguation of names and
    terms. For example, in our case study (cf.\@ \cref{sec:case_study}), we
    include a routine to deduplicate agents' names to consistently represent the
    existing ties among respondents and increase validity -- due to respondents
    who mentioned others using different monikers (e.g., nickname, surname,
    first name, full name).

    It is also important to report how heuristics can threaten the reliability
    of the data due to intrinsically personal factors -- e.g., when the
    researcher consolidates distinct references to the same person thanks to
    their knowledge of the community under study, acquired during the previous
    steps.

    \item \emph{Network analysis}: assemble network models from the respondents'
    data and apply network measures to analyse the phenomena of interest. In
    \framework{}, we both consider the network of only the respondents (a whole
    network where all nodes are equal) and the larger network that includes (the
    ties to) non-respondents (an ego-network where respondents are egos and
    non-respondents are alters).
    
    We analyse the first network under three aspects: 1) study the
    \quotes{position} (central, peripheral, isolated) of the projects'
    participants; 2) identify the more connected and central communities; 3)
    identify the people who connect communities. We reference/introduce all the
    measures in the next section.

    \item \emph{Qualitative reading}: the researcher considers all the evidence
    gathered from the qualitative and quantitative analysis of the previous
    phases and presents the emerging results. The main aim is to provide
    insights on the effects of projects on the social fabric of the studied
    community and to give directions for improving the outcomes (e.g.,
    diffusion, participation, etc.) of the current and future projects.

\end{enumerate}

\subsection{Network Measures}
\label{sec:network_measures}

We briefly detail the network measures found in \framework{}, referencing both
their original definition -- except the last measure, which is novel and
introduced in this article -- and the interpretation we give them in the context
of \framework{}. We start with the whole-network measures, followed by the
ego-network ones, including the novel one we introduce in this work.

Whole network measures are the most common ones in SNA and apply to networks
where all nodes have similar characteristics, in our case, this network
corresponds to the participants who responded to our questionnaire.

\begin{description}

\item[Core/periphery] fits a core-periphery network model to the given network,
partitioning which actors belong in either section~\citep{BE99}. Networks
showing a core-periphery structure are less decentralised but easier to coordinate.

\item[Density] calculates the ratio between the number of edges in the network
and that number if the network were complete. Social high-density networks form
redundant connections while lower-density ones likely present differentiated
relationships.

\item[Fragmentation] is the proportion of pairs of nodes that cannot reach each
other. High fragmentation values indicate the presence of sparse groups in the
network.

\item[Degree centralisation] quantifies how much degree centrality -- which
measures the importance of a node by the number of its edges -- is spread among
the nodes in a network, i.e., how much the connections are concentrated in a few
nodes~\citep{F78}. Formally, the measure is the sum of the distances of each
node's centrality from the maximal value of centrality in the network divided by
the value of that sum in a network of the same size but where centrality is
maximally spread across all nodes.

\item[Betweenness centralisation] quantifies how much the actors in the network
act as bridges~\citep{F78}. The measure's definition mirrors degree
centralisation except it replaces degree with betweenness centrality, which
measures a node's importance by its presence on the shortest paths between the
other nodes.

\item[Transitivity] measures the likelihood that two neighbours of a given node
are connected, pointing out redundancy and resilience in social ties.

\item[Modularity score] assesses how much a network breaks into
sub-communities~\citep{BGLL08}. Given a community partition of the network, the
modularity score corresponds to the distance between the number of edges within
communities and that number if the network were randomly connected.

\item[Average distance] is the average number of steps along the shortest paths
for all possible pairs of nodes. Short average distances indicate a tighter
social fabric, where all actors are close to one another;

\item[Average degree] is, in directed networks, the number of edges divided by
the number of nodes and represents the expected number of relations actors have
in the network (if evenly distributed).

\item[Assortativity coefficient] measures the tendency of nodes to others
that have similar characteristics~\citep{KS88}. In our case, we calculate
assortativity by dividing nodes \emph{by gender} and \emph{by their
participation in projects}.

\end{description}

Ego-network measures are more peculiar and apply to networks where there is a
partition of the nodes between \emph{egos} and \emph{alters}. Here, egos can
connect to any node while alters only have incoming connections from egos.

\begin{description}

\item[2-step reach] is the number of nodes an ego can reach within two steps
divided by the network's alters count. The measure indicates the size of an
ego's social influence or the reach of the ego's close social circle.

\item[Reach efficiency] is the 2-step reach divided by the number of neighbours
of the ego, quantifying how efficiently information spreads through an ego's
network -- more well-connected neighbours mean more efficiency is the spread
of information;

\item[Betweenness centrality] is, given any two nodes in the network except for
the ego, the sum of the ratio between their shortest paths that pass by the ego
and all their shortest paths. We use the normalised version of the measure,
where the sum of the centralities of the egos is one. The measure, like
betweenness centrality in whole networks, indicates the importance of the ego in
connecting its neighbourhood.

\item[Brokerage roles] computes the degree to which the ego plays five kinds of
brokerage roles. Brokerage occurs when, in a triad of nodes, two out of three
pairs are connected and one of the nodes, called \emph{source}, reaches the ego,
called \emph{broker}, and the ego reaches the other node, called \emph{sink}.
Each brokerage kind counts the number of triads where the ego is a broker and
classifies it as a
\begin{itemize}
\item \emph{coordinator}, when all three nodes belong to the same group;
\item \emph{consultant}, when the two nodes belong to the same group, different
from the one of the broker;
\item \emph{gatekeeper}, when both ego and source belong to the same group and
the sink belongs to a different one;
\item \emph{representative}, when both ego and sink belong to the same group and
the source belongs to a different one;
\item \emph{liaison}, when each node belongs to a different group.
\end{itemize}
\end{description}

\subparagraph{A Novel Measure of Importance of Alters in Ego Networks}
\label{sec:ZELDA}
2-ELDA is a novel measure, acronym of \emph{2-Ego Link Dealt by Alter}, that we
introduce to determine the influence of alters in mediating ego-ego connections.
2-ELDA overlooks directionality, focussing on the set of triplets ego-alter-ego
where the two egos have an arc to the alter and are neither connected directly
nor through another ego. 

Formally, let $G = (V, E)$ be a directed graph where $V_e \subset V$ is the set
of egos, $V_a \subset V$ is the set of alters, and $V = V_e \cup V_a$ and $V_e
\cap V_a = \emptyset$. Let $(i,j) \in E$ denote a directed edge from vertex $i$
to vertex $j$ and $P_e = \{(e_1, e_2) \mid e_1, e_2 \in V_e, e_1 \neq e_2\}$ for
all the possible ego pairs.

For any ego pair $(e_1, e_2) \in P_e$, we have direct connections $DC(e_1, e_2)
\iff (e_1, e_2) \in E \lor (e_2, e_1) \in E$, connections through another ego
$CE(e_1, e_2) \iff \exists e_3 \in V_e: ((e_1, e_3) \in E \lor (e_3, e_1) \in E)
\land ((e_2, e_3) \in E \lor (e_3, e_2) \in E)$, and connections through an
alter $CA(e_1, e_2) \iff \exists a \in V_a: (e_1, a) \in E \land (e_2, a) \in
E$.

The 2-ELDA measure is then defined as 

$$2\text{-}ELDA = \frac{|\{(e_1, e_2) \in P_e \mid CA(e_1, e_2) \land \neg
DC(e_1, e_2) \land \neg CE(e_1, e_2)\}|}{|P_e|}$$ 

Then, overloading the definition of $CA$ as $CA(e_1, e_2, a) \iff (e_1, a) \in E
\land (e_2, a) \in E$, the individual alter ranking, for each $a \in V_a$ is
$$R(a) = |\{(e_1, e_2) \in P_e \mid CA(e_1,e_2,a) \land \neg DC(e_1, e_2) \land
\neg CE(e_1, e_2)\}|$$ where $R(a)$ represents the number of (that miss both a
direct connection and one mediated by another ego) ego-ego pairs mediated by the
alter $a$.

As shown above, we propose a two-fold reading of our measure. The proper,
aggregate form, $2\text{-}ELDA$, corresponds to the number of ego couples
connected through an alter (if the egos connect via multiple alters, we count
just one) divided by the count of ego couples. Hence, the measure is the
percentage of ego couples connected through alters -- one can contextualise it
by looking at the percentage of ego couples directly connected, those connected
through another ego, and the remaining, unconnected couples. Considering the
single ego-alter-ego triplets, we can rank with \(R(a) : a \in V_a \) the alters
by their number of appearances, identifying important alters that, e.g., are
significant individuals that should participate in future study iterations. In
contexts where a participation architecture exists, 2-ELDA helps to identify
``architectural alters'' -- such as local facilitators or onboarding hubs --
that channel newcomers and coordinate action without occupying many outward ties
themselves, mirroring the tutorial/triage roles observed in networked activist
communities~\citep{MM21}.

\subparagraph*{On the Relationship between 2-ELDA, Brokerage, and Connectedness}

2-ELDA complements classical network metrics used to identify bridging positions
within organisations. Indeed, traditional measures, such as \cite{GO89}'s
brokerage roles and \cite{Burt}'s structural holes, assess how actors connect
otherwise disconnected parts of a network, focusing on the ego's own access to
non-redundant contacts or mediation of the shortest paths. To complement this
kind of measures, 2-ELDA introduces a \textit{second-order perspective},
quantifying the influence of and identifying the \textit{alter}s that connect
\textit{egos} to other egos.
 
Technically, 2-ELDA ranks alters by how many otherwise disconnected ego pairs
they bridge, revealing \textit{latent boundary partners} whose integrative
function arises from the structure itself, complementing Burt's evaluation of
redundancy in an ego's immediate ties over predefined group membership. Looking
at brokerage, 2-ELDA captures emergent bridging dynamics that rise from the
interaction between endogenous (pre-existing) organisational settings and
exogenous effects (e.g., project-based participant interactions), complementing
Gould and Fernandez's categorisation of brokerage roles. Considering cohesion,
2-ELDA complements \cite{KR94}'s concept of \textit{connectedness} at the
network level. Indeed, whereas connectedness indicates \textit{how cohesive} the
network is, 2-ELDA identifies \textit{who covertly sustains} that cohesion by
bridging ego-to-ego connections. Summarising, we see 2-ELDA as a bridge measures
between micro-level brokerage and macro-level cohesion.

In general, 2-ELDA enables a systematic detection of boundary partners -- actors
who maintain cross-unit communication, coordination, and innovation diffusion --
allowing researchers to move beyond descriptive mapping toward reproducible,
quantitative analysis of collaboration architectures.

\section{Case study}
\label{sec:case_study}

We present a case study on NGO interventions on the Pomerini village (Tanzania),
analysed through \framework{}. Since interventions aim to enhance social
cohesion, aligning with community cooperation objectives, successful projects
should induce participation relationships that make the network more cohesive.
Besides examining community cooperation interventions, the case study serves as
a meta-study to validate \framework{}. We dedicate \cref{sec:discussion} to the
meta-analysis of \framework{}'s suitability for this kind of studies.

\subsection{Context and Settings}

\begin{figure}
\includegraphics[width=\textwidth]{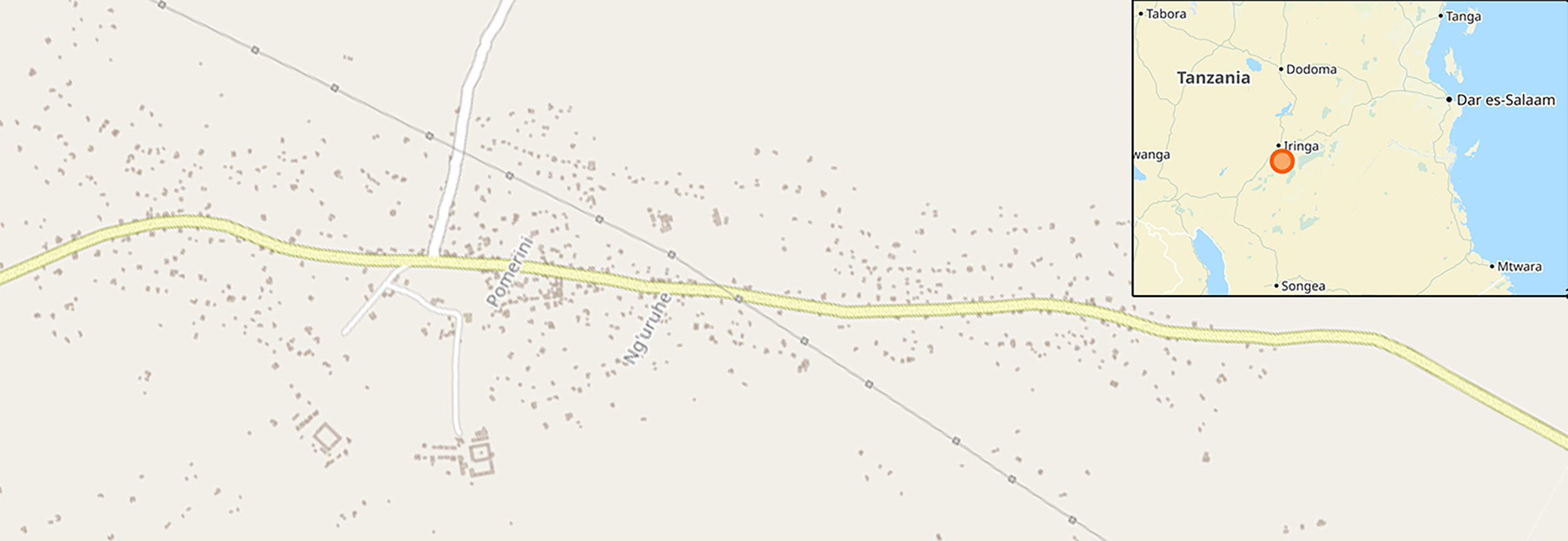}
\caption{Map view of Pomerini, Tanzania (OpenStreetMap, Dic. 2023)
\label{fig:map}.}
\end{figure}

Pomerini is a village in Tanzania's southern highlands, within Iringa's Kilolo
district. The official language of Tanzania is Swahili, but Pomerini's local
dialect is Hehe. Pomerini lacks public lighting and faces challenges with a lack
of safe running water for all households. Pomerini is part of the Ng'uruhe
territory, encompassing villages like Msengela and Msichoke. The region is
collectively referred to as ``Pomerini'' by locals due to aggregation points
therein. Ng'uruhe, with 10 streets, accommodates 13,764 people and 3,096
households, surrounded by cultivated fields. The map of the area including
buildings, in \cref{fig:map}, visualise the sparsity of the settled population.
Inhabitants travel throughout Pomerini using motorcycles but the most common
mean of moving around the area is walking, up to two hours, through villages and
fields.

The primary economic activities in Pomerini include lumbering, woodworking,
animal husbandry, clay-working, food production/distribution, and trading. Many
residents engage in multiple of these activities, often running small shops
where they sell goods like food, pottery, or household items. Some villagers
work for the government or in local public schools. Additionally, two small
dispensaries managed by local personnel operate in Pomerini.

Field research lasted 12 weeks, including four weeks for participant
observation, two weeks for defining and conducting interviews, and six weeks for
assembling, administering, and digitising questionnaires.

The field researcher collaborated with the NGO ``Tulime'' to organise the field
study. Tulime provided accommodation in ``Casa Tulime'', situated between
Pomerini and Msengela, near a Franciscan mission and the Roman Catholic Church.
The researcher conducted the study activities personally, assisted by a local
facilitator, \prt{S}{N}, proficient in both English and Swahili.

All participants provided informed consent for the non-commercial use of their
anonymised data. School headmasters granted permission for the use of anonymised
data of students under their administration.

\subsection{Phase 1: Participant Observation}

\prt{S}{N}, the cultural mediator, facilitated the researcher's involvement in
project activities in Pomerini, and also suggested strategies to observe
different aspects of village life and connect with potential key informants.

Beyond Tulime, other NGOs operate in Pomerini. The Franciscan mission hosts the
NGO ``Mawaki'', which legally owns Casa Tulime. Casa Tulime is also shared with
a local family of four people, a couple, \prt{N}{X} and \prt{M}{Y}, and two
children. \prt{N}{X} works for the nearby Agriculture University, which is also
under the management of Mawaki. Furthermore, a lady is working in the house
during volunteers or researchers' stay, \prt{A}{K} -- who was first a
beneficiary of Mawaki's microcredit, allowing her to provide superior
instruction to her daughter, \prt{N}{M}, and later a beneficiary of Tulime's
microcredit, which she invested on her farm. \prt{A}{K} is in charge of cooking
during the week and taking care of the house before volunteers arrive and after
they leave. \prt{N}{M} is the current headmaster of Mawaki's Agriculture
University. The Tabasamu centre, close to the mission, houses two more NGOs,
``Smile to Africa'' and ``Nyumba Ali'', along with a dispensary managed by
monks. These NGOs support children with disabilities or HIV, providing medical
care and daily physiotherapy. Mawaki, in addition to offering cars for
emergencies, lends agricultural machinery to local farmers. Both Tulime and
Mawaki started operating in Pomerini in the early 2000s, while Smile to Africa
and Nyumba Ali joined in the following decade.

Most villagers have access to public water, thanks to the fountains provided
through the ``Maji'' project. Nonetheless, both the mission and \prt{A}{K}'s
house have wells used by the villagers. The ``Maji'' project is now under the
care of village committees. Among the villages of the area, Kitowo decided in
2019 to entrust Mawaki, which managed the installation of wells and pumps and
was in charge of water management until the funds ended; during the research
period, the village was electing a proper committee, as required by Tanzanian
laws.

Mawaki has a wider set of projects, focused mainly on microcredit, schooling,
water projects, and agriculture interventions. Tulime's projects, many of
which intertwine with the ones performed by Mawaki and Smile to Africa,
include microcredit, water projects in five different villages (``Maji''), a
pre-kindergarten (``Tupo Pamoja''), a cyclo-mechanical workshop
(``Baiskeli''), a reforestation and wood production project affiliated with
schools (``Plant it now!''), a tailoring workshop (``Hands of Africa''), a
university scholarship for students (``Mwalimu''), and a workshop for disabled
women created along with Nyumba Ali (``Be Able''). In the past years, there
was also a project to enforce good environment preservation and recycling
practices called ``Environmental Club Network'', preceding the school projects. 

Several schools surround Pomerini, operating under the Tanzanian public school
system. Pomerini's Primary School, Msengela's Primary School, and Kihesa Mgagao
Institute are part of Tulime's ``Plant it now!'' project. Pomerini's Primary
School accommodates 567 students, seven classes, and 21 teachers, Msengela's
Primary School has 420 students, eight classes (encompassing all primary years
and one pre-primary), and nine teachers, and The Kihesa Mgagao Institute counts
436 students, eight classes (encompassing all primary years and one pre-
primary), and nine teachers. Additionally, 97 students with disabilities reside
in the school building, receiving care and education. The school is equipped
with two small laboratories for auditory and visual impairments, along with the
materials to build and repair hearing aids. Thirteen of the 21 teachers working
in the school have preparation for dealing with the mental and physical
disabilities of students, and they all speak Swahili Sign Language to
communicate children who are deaf. Four teachers manage Tupo Pamoja (the
pre-kindergarten under Tulime's care in Pomerini), which hosts 32 children,
alternating in groups of two. There is also a kindergarten in Mawaki (in the
mission's backyard), hosting 139 children, divided into 3 classes, and managed
by 3 local teachers and some monks. The children in Tupo Pamoja can use a
cyclo-bus service to get to school and back home every day, consisting of a
small cart attached to a bike. The drivers are Tulime's beneficiaries and
Baiskeli's workshop realised the cyclo-buses. All the teachers of Tupo Pamoja
were initially involved in other Tulime's activities: \prt{C}{K} and \prt{E}{C}
as micro-credit beneficiaries, \prt{M}{M} as a ``Be Able'' operator  --
\prt{M}{C} was involved as a relative of \prt{E}{C}.

Before the travel restrictions due to the COVID-19 pandemic, volunteers often
stayed at Casa Tulime, contributing to various projects. However, from 2020 to
the summer of 2022, no one resided in the building. The projects continued with
local participation and organisers, with Smile to Africa and Nyumba Ali's
activities carried out by locals. In contrast, Mawaki, maintained a stable
presence thanks to its affiliation with the Franciscan mission.

\subsection{Phase 2: Semi-structured Interviews}

Our data collection phase amounted to 12 weeks. All its components were
carried out with the help of a local facilitator, \prt{S}{N}, a young lady who
speaks both English and Swahili and who previously collaborated with Tulime's
volunteers on several occasions. The information gathered in the previous step 
allowed us to identify seven key informants to interview. We selected the 
informants from among Tulime or Mawaki project beneficiaries or local collaborators.

The design of the interview structure encompassed three steps. First, the
researcher developed the draft in English, focussing on questions useful
to clarify obscure elements from the observation phase, such as how locals spent
leisure time or family dynamics in the village (which would require a longer
observation phase). Second, the researcher discussed the questions with the
cultural mediator. Interventions on the initial draft regarded the rephrasing of
sentences that would poorly apply to the context of the study (e.g., the initial
distinction between family and cohabitation ties) or which would generate
ambiguous translations in Swahili. The third step included the translation and
adjustment of the questions to align the meaning of the English and Swahili
versions and checking for consistency and adherence to our informative needs. At
the end of this process, almost all questions had specific subquestions to
clarify their intended meaning.

The researcher administered the interviews accompanied by \prt{S}{N}. The researcher
conducted and transcribed the interviews with proficient English speakers.
\prt{S}{N} conducted the interviews in Swahili; she also translated the
responses into English and the researcher transcribed them.

A common trait that emerged from the interviews is that people participated in
mostly one project, with a few exceptions, e.g., \prt{E}{C}, who was a
beneficiary of the Microcredit project, a teacher in Tupo Pamoja, and
responsible for (and involved in) all the active projects or \prt{R}{K}, former
microcredit beneficiary and owner of the tailoring shop. Both \prt{E}{C} and
\prt{R}{K} autonomously identified one activity as their reference project (Tupo
Pamoja and Hands of Africa, respectively), while the other activities came up
during their interviews. In general, all people who took part in more than one
project pointed out a specific one autonomously while the others emerged as the
interviewer inquired about their lives in increasing depth.

\subsection{Phase 3: Administration of Questionnaires}
The interviews influenced the development of the questionnaire (e.g., on
bounding the number of family members), outlining its structure with 19
questions about respondents' relationships. These types of relations include
``Family'', ``Friend'', ``Co-worker'', and ``Others'' -- the latter are
important relations that do not fall into the other categories, e.g.,
mentorship. Moreover, we asked the respondents to indicate relationships with
people they met thanks to project participation, labelling these relations as
``Relation from Projects'' (these relations can intersect the other ones, e.g.,
Friends and Co-workers) and people involved in projects that they knew before
their participation, labelled ``Pre-existing Relations''.

Questions also addressed the impact of COVID-19 on Tulime's presence in the
village, aiming to understand how external events could alter the social fabric
quantitatively. Since our main focus is on the validation of \framework{}, we only
consider the data related to the post-pandemic situation of the village. Nonetheless, 
further analyses of the data also provided us with a measure of the resilience of 
Pomerini's social fabric to exogenous shocks (i.e., COVID-19): this topic will not be
discussed in this article, given its extremely low relevance to the central topic
of this study (namely the presentation and validation of our framework). In general,
we verified that the pre- and post-pandemic configurations of the dataset do not
substantially vary, with the main difference regarding 5 international
cooperators who did not return to Pomerini.

The researcher administered the questionnaires face-to-face over 6 weeks to
Tulime's project participants, students, Mawaki's staff, mission monks,
kindergarten teachers, and Tabasamu nurses. Primary school staff was not
available, but headmasters were briefly interviewed to qualitatively assess
project impacts. The total number of respondents is 382.

\subsection{Phase 4: Digitisation and Data Cleaning}

We digitised pen-and-paper questionnaire responses in weekly batches (hence,
this step and the previous one alternated over the 6 weeks of the questionnaire
administration), checking the correspondence between the digital version and its
source to minimise transcription errors.

Analysing the responses, we found missing or ambiguous information. Thus, we
devised a semi-automatic data-cleaning process to apply study-wide but
case-by-case intervention on the data, supported by the knowledge acquired in
the previous phase. While the process is standardised, the specific
interventions introduce a trait of subjectivity that threatens reliability. To
mitigate this threat, the interventions were first proposed and explained by the
in-field researcher to at least one of the other authors. The latter approved
the proposals only after they had enough information to reasonably inform their
decision e.g., consulting the field journals.

Corrections included: spelling mistakes in names (e.g., Hadija/Adija,
Emanuel/Emanueli), reconciling to a name-surname format the different monikers
used by the respondents, and disambiguation of homonymy.

Homonymy cases mainly regard students, which we addressed by assuming they chose
desk mates based on spatial closeness. We inferred the spatial relations from
the questionnaires' order, which were gathered to mirror the position of the
students in classes. For surname attribution, mainly in polygamous families, we
paired families with the same surname as the respondent's answer. We
consolidated entries related to family members based on spelling suggestions
from other respondents, aligning entries with partial or complete equivalence.

For reference, we report in \cref{tab:relationships} the number of respondents
with at least one tie of that relation type and the percentage of them with at
least one modified tie of that relation type -- on average, we fixed 1--5 entries
for each response we modified. We note that the substantial modifications
concentrate on friendship and co-work relations, which we attribute to the use
of nicknames. We moved the rows of ``Relations from Projects'' and
``Pre-existing Relations'' at the bottom of the table because these types can
include the other types of relations and their percentage of modifications
results as an average of the other relation types. The percentages do not
include straightforward modifications like name spelling corrections.

\begin{table}[h]
    \centering
    \begin{tabular}{lccc}
      \toprule
      Relation type & \makecell{Number of respondents \\ with at least one tie\\
      of that relation type} & \makecell{Percentage of respondents\\ with at
      least one modified tie\\ of that relation type}\\
      \midrule
      Friends                 & 375 & 63.5\% \\ Co-workers              & 62 & 40.3\% \\ Others                  & 187 & 35.8\% \\ Family                  & 381 & 28.6\% \\ \midrule
      Relations from Projects & 204 & 51.5\% \\ Pre-existing Relations  & 129 & 37.2\% \\
\bottomrule
    \end{tabular}
    \caption{Table reporting the number of respondents with at least one tie of
    that relation type and the percentage of them with at least one modified tie
    of that relation type.}
    \label{tab:relationships}
  \end{table}

\subsection{Phase 5: Network Analysis}

As mentioned in \cref{sec:framework}, \framework{} comprises the analysis of two
networks. One is the ego-network that represents the respondents and their
connections to others who might not be respondents themselves. The other is a
whole-network obtained by removing non-respondents from the first one.

The ego-network, from now on \(G_e\), includes 2881 nodes and 5222 edges. The
whole-network, dubbed \(G_w\), includes 382 nodes and 1885 edges. The edges in
both networks carry their labelled type, as found in \cref{tab:relationships}.

In \(G_e\), we observe the presence of 5 small isolated components ranging five
to eleven nodes, for a total of 41 nodes (1.42\%) and 36 edges (0.75\%). Since
these components are isolated and small, we reasonably focus the rest of our
analysis on the main component and remove these nodes and related edges from
both \(G_e\) and \(G_w\).

To study the influence of projects on the social fabric of the community, we
consider \(G_e^p\) and \(G_w^p\) as variants of resp.\@  \(G_e\) and \(G_w\)
where we remove the relations induced by project participation.

\subsection{Phase 6: Qualitative Reading}

We applied the measures presented in \cref{sec:framework} on the gathered
networks. Following the partition in \cref{sec:framework}, we start from the
results on the whole-network and then pass to analyse the ego-network.

\subsubsection{Whole-network Measures}

\begin{table}[t]
\centering
\begin{tabular}{p{0.3\linewidth}|p{0.3\linewidth}|p{0.3\linewidth}} & \(G_w\) &
    \(G_w^p\)\\
    \hline
    \hline
    Core/periphery & 0.5443 (20 iterations) & 0.6355 (20 iterations) \\
    \hline
    Density & 0.009 & 0.009 \\
    \hline
    Fragmentation & 0.42 & 0.454 \\
\hline
    Betweenness centralisation & 6.42\% & 7.59\% \\
    \hline
    Transitivity & 0.473 & 0.495 \\
    \hline
    Degree centralisation & 0.057 (out 0.046, in 0.054) & 0.055 (out 0.044, in
    0.044) \\
    \hline
    Modularity score & 0.784 (30 communities) & 0.764 (38 communities) \\
    \hline
    Average distance & 7.995 & 8.529 \\
    \hline
    Assortativity by gender & 0.802 & 0.796 \\ 
    \hline
    Assortativity by projects & 0.698 & 0.701 \\
    \hline
    Average degree & 3.531 & 3.28 \\
    \hline
\end{tabular}
\caption{Whole-network measures, comparing the whole-network \(G_w\) against the whole-network without the relationships generated from projects' participation \(G_w^p\).}
\label{tab:measures_table}
\end{table}

\newcommand{\lgdsq}[4]{\raisebox{.25em}{\color[RGB]{#1}\mbox{\rule{.05\textwidth}{0.3mm}}}
& \mbox{#2} & \mbox{(#3\%/#4\%)}} \newcommand{\lgddot}[3]{
  \definecolor{tmpcolr}{RGB}{#1}
  \tikz\draw[black,fill=tmpcolr] (0,0) circle (.6ex); & \mbox{#2} &
\mbox{(#3\%)} }

\begin{figure}[t]
\centering
\includegraphics[width=\linewidth]{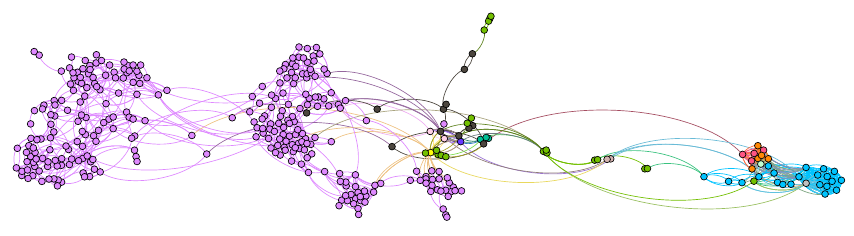}
\includegraphics[width=\linewidth]{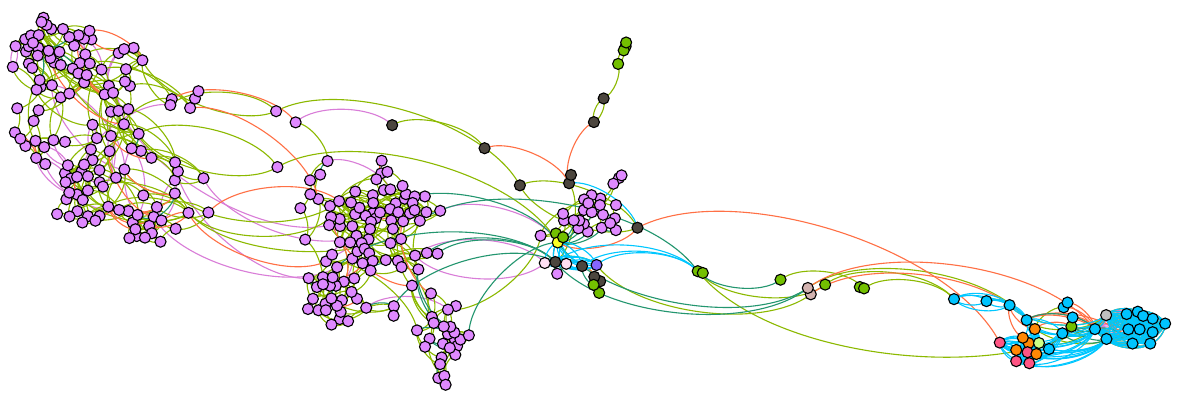}
\vspace{1em}
\begin{adjustbox}{width=\textwidth}
\(\begin{array}{cllcllcll} \\
\lgdsq{30,201,0}{Friend}{50.56}{61.38}
& \lgddot{237,132,255}{Plant it now!}{77.23} & \lgddot{255,65,131}{Hands of
Africa}{0.79} \\ \lgdsq{250,127,0}{From Projects}{17.63}{-} &
\lgddot{85,194,0}{Microcredit}{8.64} & \lgddot{217,178,175}{Mwalimu}{0.52} \\
\lgdsq{0,206,255}{Work}{15.45}{18.76} & \lgddot{0,200,255}{Mawaki}{5.5} &
\lgddot{255,207,231}{Baiskeli}{0.52} \\
\lgdsq{209,130,239}{Family}{10.04}{12.19} & \lgddot{77,70,61}{Tupo Pamoja}{3.4}
& \lgddot{176,255,207}{Tabasamu}{0.26} \\
\lgdsq{0,146,112}{Other}{3.08}{3.74} & \lgddot{255,128,0}{Tembea}{1.31} &
\lgddot{109,53,255}{BeAble}{0.26} \\ \lgdsq{255,75,127}{Pre-existing
Rel.}{3.24}{3.93} & \lgddot{85,194,0}{Smile to Africa}{1.05} &
\lgddot{255,253,0}{Local proj. resp.}{0.26} \end{array}\)
\end{adjustbox}
\caption{Top \(G_e\), bottom \(G_e^p\). Force Atlas 2 \citep{Jacomy2014} layout
with colour-coded nodes (project participation) and edges (relationship) with
percentages (of \(G_w\) on the left and of \(G_w^p\) on the right).
\label{fig:whole_networks}}
\end{figure}

We report in \cref{tab:measures_table} the set of whole-network measures on both
\(G_w\) and \(G_w^p\) and visualise the networks in \cref{fig:whole_networks}.

Starting from the top of \cref{tab:measures_table}, we note that Core/periphery,
Density, and Fragmentation indicate a substantial level of centralisation of
\(G_w\), along with a low density of ties and a relatively high fragmentation.
The connotation is more accentuated for \(G_w^p\), indicating that projects
decentralise the community -- instead of relying on a few central figures to
connect the network. The phenomenon emerges also in \cref{fig:whole_networks},
where links among different communities created by participants in international
cooperators' projects. Removing project-induced ties increases core-periphery
contrast, average distance, fragmentation, and betweenness centralisation
(\cref{tab:relationships}). This result is consistent with participation
architecture predictions, since project routines appear to decentralise
brokerage and shorten paths by introducing many small portals for collaboration
rather than funnelling coordination through a few hubs (cf.~\cref{sec:state_of_the_art}).

Betweenness centralisation, Transitivity, Degree centralisation, Modularity
score, Average distance, by-gender and by-projects Assortativity point to a
tendency to ease information flow and a loose structure, still concentrated
around more central communities able to dominate the network, with a propensity
to build relationships with people of the same gender and belonging to the same
projects, and a generally high distance between participants. Homophily values
show little variations between \(G_w\) and \(G_w^p\). We interpret this result
as an indication that projects develop over existing relationships, although
participation in projects creates new heterophilic connections.

The increase of these values from \(G_w\) to \(G_w^p\) confirms the positive
action of projects in making the network less centralised and more connected.
Looking at \cref{fig:whole_networks}, we can refine the assertion above by
identifying a stronger division between the communities in \(G_w^p\), which can
lead to information/resource bottlenecks due to the few nodes that serve as
bridges between the tighter conglomerates.

The values of Average degree, Density, Degree centralisation, and Fragmentation
provide a reading of the people able to connect communities. The values reveal
that connections are sparse and fragmented in \(G_w\), but the situation is
worse for \(G_w^p\), where connections are fewer and even more fragmented.
Looking at \cref{fig:whole_networks}, we notice in particular fewer connections
from the Microcredit beneficiaries. Those respondents, in fact, mainly act as bridges
between communities but do not present many ties with other participants.

\subsubsection{Ego-network Measures}

\begin{table}[t]
\centering
\begin{adjustbox}{width=\textwidth}
\begin{tabular}{c|c|c|c|c|c|c}
  & \prt{E}{C} & \prto{Brother }{P.} & \prt{G}{M} & \prt{A}{K} & \prt{C}{K} &
  \prt{R}{K} \\
  \hline
  Density & 18.18 | 21.11 & 4.58 | 4.90 & 10.80 | 11.58 & 0 | 0 & 10.48 | 10.48
  & 7.19 | 1.28 \\
  \hline
  Average distance & 0.20 | 0.22 & 0.09 | 0.10 & 0.13 | 0.14 & 0 | 0 & 0.11 |
  0.11 & 0.07 | 0.01 \\
  \hline
  Two-step reach & 64 | 62 & 251 | 222 & 107 | 101 & 50 | 45 & 102 |
  95\phantom{0} & 128 | 51\phantom{x} \\
  \hline
  Reach efficiency & 61.54 | 64.58 & 27.29 | 27.00 & 34.74 | 33.67 & 98.04 |
  100.0 & 69.39 | 70.37 & 73.14 | 94.44 \\
  \hline
  \makecell{Betweenness \\centrality\\ (normalized)}  & 16.14 | 18.06 & 7.12 |
  8.06 & 13.79 | 13.79 & 6.67 | 6.67 & 15.68 | 15.68 & 10.46 | 14.10 \\
  \hline
\end{tabular}
\end{adjustbox}
\caption{Ego-network measures, comparing the ego-network \(G_e\) (values on the
left in cells) against the ego-network without the relationships generated from
project's participations \(G_e^p\) (values on the right in cells).}
\label{tab:ego_measures_table}
\end{table}

Since the measures for the ego network regard nodes (contrarily to whole-network
measures), we report in \cref{tab:ego_measures_table} the ego-network measures
computed on the nodes that emerged as more meaningful from the observation and
interview phases (i.e., phases 1 and 2). We choose to present the data in this
way both for brevity and because many respondents belong to the schools of
Pomerini, Msengela, and Kihesa Mgagao and show relationships mainly with fellow
students, apart from rare exceptions, that are less relevant in this analysis.
In \cref{tab:ego_measures_table}, we report the values of each measure for
\(G_e\) and \(G_e^p\) resp.\@ on the left and right side of each cell.

To select the nodes, we consider those who have management roles or emerged as
involved in the life of the village (typically the two traits overlap). We note
that, while \prt{E}{C}, \prto{Brother }{P.}, and \prt{G}{M} manage a high number
of people, \prt{C}{K} and \prt{R}{K} are in charge of smaller projects,
involving 2 to 3 people, with whom they collaborate rather than directing them.
\prt{A}{K} is not in charge of any project, but she showed a deep knowledge of
other Microcredit beneficiaries and is in charge of the management of Casa
Tulime.

In both networks, \prt{E}{C} exhibits higher density values compared to
\prt{G}{M} and \prt{C}{K}. Following them are \prto{Brother }{P.} and
\prt{R}{K}, with the latter displaying higher values in the network representing
the current community. The average distance shows a similar pattern, except
\prt{R}{K} consistently has the lowest value. Notably, \prt{A}{K} has a value of
0 for both measures, as she does not connect any pair of alters, possibly due to
her diverse relations directed towards individuals often not mentioned in the
network.

The dense connections of \prt{E}{C} suggest her ability to form tight-knit
communities around herself. Interestingly, her density score is higher in the
network excluding project ties, indicating that her role as a local responsible
links her to different communities, expanding the number of her alters and
reducing her density. \prto{Brother }{P.} has a broader network that involves
co-workers and friends. Moreover, he holds a central role in Mawaki and serves
as a general figurehead in the village due to his engagement in local projects
and interventions.

Conversely, \prt{R}{K} manages a smaller network but, as the project manager for
a local tailoring shop, her alter community appears smaller, tighter, and
positively influenced by project ties. Both \prt{C}{K} and \prt{G}{M} exhibit
similarly low density and average distance values. We attribute this trait to
their management of small projects centred around close-knit communities of
co-workers and collaborators, resulting in low distances among their alters and
high-density values for themselves.

\prto{Brother }{P.} has the highest two-step reach values, while \prt{A}{K} has
the lowest. Conversely, \prt{A}{K} exhibits the highest reach efficiency, with
\prto{Brother }{P.} having the lowest score. In the network, the community
around \prto{Brother }{P.} is broader but more redundant compared to the one
involving \prt{A}{K} and the relations from projects (\(G_e^p\)) seem to have a
low impact on their communities.

\prt{C}{K}'s two-step reach shows smaller variations, indicating a stable
community before project participation. \prt{R}{K}'s two-step reach is more
affected, suggesting that her community consists of small sets of well-connected
acquaintances. 
\prt{G}{M}, managing the Tabasamu centre, exhibits a pattern akin to \prt{C}{K}.
In contrast, from the network, \prt{E}{C} has a cohesive yet open group before
projects. As the local responsible for managing projects in Pomerini and
Msengela, her consistent values suggest the people involved in her projects are
already part of her community. These observations align with the obtained
betweenness values.

Considering the brokerage roles we find other important nodes. Specifically,
\prt{E}{C} is a Liaison, fitting with her role as the local projects'
responsible. \prt{C}{K} and \prt{M}{C}, the two teachers from Tupo Pamoja, are
Liaisons in all networks and they connect the group with others, while, among
the mothers, \prt{M}{U}, \prt{F}{M}, and \prt{C}{K} are Gatekeepers, \prt{V}{M}
is a Coordinator and \prt{D}{M} is a Representative. The Coordinators that
emerge in the network excluding projects' ties indicate that relations between
the mothers involved in Tupo Pamoja existed before the project started.
Analysing the Microcredit beneficiaries, we note that, in \(G_e\), \prt{A}{K},
\prt{S}{M}, and \prt{L}{T} are Liaisons, while \prt{J}{K} is a Gatekeeper and
\prt{M}{M} and \prt{E}{M} are Representatives. When observing \(G_e^p\),
\prt{L}{T} becomes the only Liaison. Since Microcredit does not exist in
\(G_e^p\), we infer that \prt{L}{T} was connecting the micro-creditors with
people external to the projects. Since \prt{L}{T} is the wife of \prt{E}{M} --
one of the central figures of Mawaki -- she mostly had relations with people
from Mawaki. \prt{A}{K} assumes the role of Gatekeeper, underlining her role as
a trusted figure in the village, while \prt{V}{T} -- the market seller -- and
\prt{J}{K} -- the pharmacy owner -- become Coordinators, underlining their
existing relations with many of the people involved in the projects.

Finally, considering the brokerage roles involving Mawaki, both in \(G_e\) and
\(G_e^p\), we observe that most of the monks work as Coordinators, thus
connecting other members of their community. From our observations, apart from
Mawaki, monks live and work together in the mission, mostly developing working
and ``family'' ties with one another. In \(G_e^p\), we find \prt{E}{M} and
\prt{K}{M} -- the two drivers also working with Tulime -- as Liaisons who
connect others to the community.

In \(G_e^p\), we find that \prto{Brother }{P.} and {Brother }{S.} have a
Representative role and connect the community with other people. We explain this
fact from collaborations with other projects, such as the Tabasamu centre, which
pass through their relations with other projects' representatives.

\subsubsection{2-Ego Link Dealt by an Alter} 

Applying our novel measure on \(G_e\), we observe 976 direct connections
among nodes, 2950 distinct step-2 connections among egos, 248 Ego-Alter-Ego
connections and 237 distinct connections, involving a total of 186 participants.
The disconnected ego couples are 68608 and the connections mediated by alters
(237), are 0.34\% of all possible connections. The top 10 alters mediating
connections among egos are \prt{S}{N} (59), \prt{H}{I} (9), \prt{P}{C} (7),
\prt{S}{C} (7), \prt{O}{K} (5), \prt{E}{M} (5), \prt{G}{M} (5), \prt{Y}{N} (5),
\prt{R}{C} (4), \prt{S}{K} (3).

Applying 2-ELDA on \(G_e^p\), we observe 1059 direct connections among
nodes, 3441 distinct step-2 connections among egos, 351 Ego-Alter-Ego
connections and 333 distinct connections, involving a total of 195 participants.
The disconnected ego couples are 67938 and the connections mediated by alters
(333), are 0.49\% of all possible connections. The top 10 alters mediating
connections among egos are \prt{S}{C} (83), \prt{S}{N} (78), \prt{H}{I} (9),
\prt{Y}{N} (8), \prt{F}{P}  (7), \prt{P}{C} (7), \prt{G}{M} (5), \prt{E}{M} (5),
\prt{O}{K} (5), \prt{R} (5).

The number of connections passing through cooperators amounts to 7 in \(G_e^p\),
and 95 in \(G_e\). If we also consider \prt{S}{N}, the local facilitator, they
amount to 66 and 173, respectively. \prt{S}{N} took part in Tulime's and
Mawaki's projects until 2019, and she knew most of the people in the village, in
particular Microcredit beneficiaries, to whom she delivered pay cheques when the
project was still ongoing, but she was not directly involved in the study
because she is not involved in any project nor does she perceive any
compensation from Tulime.

Besides \prt{S}{N}, the top 2-ELDA ego is \prt{S}{C}, who is Tulime projects'
responsible. This denotes that i) relations are by and towards international
cooperators before proper project development, thus suggesting a real commitment
towards Community Cooperation from the initial phases of project ideation; ii)
in some cases, international cooperators used the local facilitator to establish
a connection with individuals belonging to different communities in the village,
and iii) some ``peripheral'' nodes/participants, who connected to a given
community through project participation, lost their connection to other members
of that community when international cooperators and \prt{S}{N} disappeared due
to COVID-19.

Finally, \prt{H}{I}, \prt{R}{K}'s husband, also appears as a bridge for 9
connections in the network excluding ties related to projects, and 5 ties in the
network excluding relations lost since 2020. In both cases,
he serves to connect \prt{T}{C} and \prt{D}{M} (and in the first case also
\prt{R}{K}) to other Microcredit beneficiaries and people involved with Mawaki.

\subsubsection{Threats to Validity}

The first two phases of \framework{} (participant observation and interviews)
intrinsically require actions and decisions taken by the researcher, who can
introduce their bias in the interpretation of the community and the
interviews, as well as in choosing which people to integrate in the study.

Questionnaires allow us to gather first-hand data that describes the studied
context. However, we recognise (and quantify) the heterogeneity in the accuracy
of the answers by the participants, which requires a cleaning procedure where
the researcher's perspective helped to consolidate the data. This step
introduces a threat to the reliability of the process and the validity of the
dataset, due to the inherent influence of the perspective of the researcher.

While the step of network analysis does not introduce relevant threats to
validity, the qualitative reading of the results (mixing the quantitative
results from the measures with the qualitative observations of the researcher)
presents the same bias that characterises the first phases. 
\section{Discussion and Conclusion}
\label{sec:discussion}
We conclude by discussing the results of the case study both regarding the
effect of NGO projects on Pomerini's social fabric and from a meta-analysis
point of view on the usage of \framework{} for this kind of studies. Moreover,
we complete the positioning of our novel measure, 2-ELDA, within the context of
organisational theory and network science and discuss this study's limitations
and comment on future evolutions.

\subsection{Remarks on the Case Study}
The results from the case study reveal that participants are already engaged in
a broad social pool or maintain largely unchanged close relationships regardless
of their project participation -- where adults have slightly wider and more
diversified personal networks thanks to project ties.

Nonetheless, we found positive effects from projects, lowering network
centralisation and fostering community aggregation, although these communities
appeared disconnected, with few nodes serving as bridges. Many nodes connecting
isolated individuals to the broader community are international cooperators,
most notably for micro-creditors, whom the former connect with the rest of the
network.

The findings stress the importance of creating new, cohesive connections between
participants, emphasising the importance of fostering ties beyond pre-existing
relationships for a project's success.

For example, \prt{S}{N} (local facilitator) and \prt{S}{C} (Tulime projects'
responsible) emerge as top 2-ELDA alters, i.e., nodes that mediate otherwise
disconnected ego pairs. Substantively, these actors look like “architectural
alters”: their roles resemble the onboarding/tutorial gateways seen in
participation architectures, which unobtrusively channel newcomers and route
coordination across clusters. The COVID-19 withdrawal of international
cooperators (and, intermittently, facilitators) helps explain why several
ego-alter-ego conduits collapsed, mirroring participation bottlenecks described
when architectural gateways recede. This evidence reinforces an important trait
of successful project design: institutionalising local architectural roles so
that coordination capacity persists when external staff cycles.

\subsection{Remarks on the Research Framework}

The case study allows us to draw practical remarks on \framework{}.

The anthropological analysis of the study setting plays a crucial role in the
research's development. Field observation enables a fundamental understanding of
village dynamics, project functioning, and the identification of key informants
for interviews. The time invested in observing the local population fosters
familiarity and trust, positively influencing the conduction of interviews and
questionnaires.

Measure-wise, we note some redundancies. For example, Core/periphery and Degree
Centralisation both convey the network's structure, and Density and Average
Degree reflect its connectivity. Although slightly redundant, we deem it
useful to perform all measures to confirm the consistency of the results.

Moreover, our interpretation of the results suggests improving the collaborative
design of projects with the broader local community, preventing international
cooperators from becoming the sole link between participants, adverted by
involving local representatives as active members of interventions. Indeed,
participants demonstrated commitment and cooperation with external
collaborators, suggesting the possibility of leveraging their presence to
enhance the network's ability to integrate temporary collaborators.

Similarly, the results advocate for synergic cooperation among associations.
Operators and representatives across different projects should coordinate to
establish tightly-knit connections among themselves and other ``central''
participants who can extend the overview and reach of projects, in particular
for what regards people such as \prt{S}{N} and \prt{H}{I}, who emerged as
significant 2-ELDA alters.

Read through the participation-architecture lens, our results suggest that NGO
projects in Pomerini did not simply ``add ties''; they configured a lightweight
architecture that (while not codified as a points system) multiplied entry
points and distributed brokerage. The reliance on a handful of architectural
alters (\prt{S}{N}, \prt{S}{C}, and a few international cooperators) implies
fragility when these gateways disappear. Designing simple, fair, and transparent
involvement and delegation routines for locals would increase the degree of
embedding of in-situ architectural control and further reduce centralisation and
fragmentation.

\subsection{Contributions to Organisational Theory and Network Science}
Interpreted through organisational theory, the high 2-ELDA alters that emerge
from the study correspond to \emph{boundary-spanning} actors who bridge
otherwise segregated groups and knowledge domains \citep{TU81,LE5}. Classic
brokerage perspectives emphasise how actors connect disconnected others, either
via roles across categorical boundaries \citep{GO89} or by occupying structural
holes that provide access to non-redundant information \citep{Burt}. Our use of
2-ELDA complements these approaches by focusing on the \emph{alter's}
second-order mediation. The measure identifies ``boundary partners'' who link
ego-ego pairs that are neither directly connected nor joined through other egos.
In this sense, architectural alters, in Pomerini, function as \emph{tertius
iungens} catalysts that weave ties across clusters \citep{OB5}. At the
organisational level, these agents serve as enablers for cross-unit coordination
and the diffusion of practices \citep{CU5,CA2}.

At the network level, the observed reduction in centralisation, accompanied by a
reliance on a small set of architectural alters, suggests a tension between
emergent cohesion and fragility. This result dovetails with whole-network
measures, such as \emph{connectedness} \citep{KR94}, whereby, when overall
cohesion improves, the concentration of bridging capacity among a few gateways
implies vulnerability if these actors exit. Designing for \emph{local}
boundary-spanning capacity -- i.e., employing 2-ELDA scores to embed external
cooperators in recognised community roles -- aligns qualitative design choices
with quantitative expectations of lower betweenness centralisation, weaker
core-periphery structure, shorter paths, and more distributed brokerage.

Methodologically, \framework{} adds depth to the analysis of boundary 
spanners by integrating 2-ELDA with standard metrics. Indeed, whereas
brokerage and structural-hole measures typically assess an actor's
\emph{egocentric} access or their presence on shortest
paths~\citep{GO89,Burt,F77}, 2-ELDA quantifies an \emph{alter's} second-order
mediation of ego-ego pairs which lack direct or ego-mediated ties. This
observation leads us to deem 2-ELDA well-suited for detecting emergent,
exogenous-based (e.g., via projects) settings where formal group boundaries are
fluid, and complements whole-network cohesion measures, like
connectedness~\citep{KR94}. In longitudinal replications, pairing 2-ELDA with
centralisation, core-periphery, and path-length statistics can separately track
(i) changes in overall cohesion and (ii) the distribution of boundary-spanning
capacity across local vs.\ external actors.

\subsection{Limitations and Future Work}
While the application of \framework{} and the 2-ELDA measure provides new
insights into the architecture of collaboration, several limitations remain.

First, we used \framework{} in a study that captures a single temporal snapshot
of the studied network dynamics; hence, causal inferences about how
boundary-spanning capacity evolves over time remain tentative. Moreover,
networks related to the social fabric before interventions and before COVID-19
have been reconstructed by participants, not directly observed, most likely
introducing errors and biases (e.g., omission errors due to memory and/or
interpretation of past social trends in light of the current situation).

In general, given a wider timespan and the appropriate settings, we suggest
repeating the same study at different moments (mainly after the
deployment/conclusion of interventions), which would provide even more
insightful results on the evolution of the analysed social fabric. This item
represents interesting future work. Essentially, it entails the refinement of
\framework{} to include meta-measures that integrate data from past studies,
allowing for both the quantification and visualisation of the evolution of the
network and supporting the conduction of what-if analysis and predictions of the
network's evolution.

As a meta-contribution, \framework{} offers a way to measure the presence and
effects of participation architectures in development settings. Future
replications could code architectural features during project design and initial
phases, and then test whether they predict lower betweenness centralisation,
lower core-periphery, shorter average distances, and higher 2-ELDA scores for
local (vs external) alters. These results would align the qualitative design of
interventions with quantitative network outcomes, providing both a qualitative
reading and a potentially predictive function to our framework. Longitudinal
data would support, for instance, to explore whether increases in local 2-ELDA
scores precede improvements in cohesion or project resilience, thus testing the
hypothesised link between distributed brokerage and adaptive capacity. 

Second, although 2-ELDA identifies alters mediating ego pairs, it does not yet
incorporate directionality or tie strength, potentially overlooking asymmetries
in influence and communication flow. Future extensions could weight
ego-alter-ego triplets by interaction frequency or reciprocity, aligning the
measure more closely with theories of relational embeddedness~\citep{GV91}.

From the point of view of the conducted case study, we have limited
generalisability due to the context-specific nature of NGO collaborations.
Applying \framework{} to other settings -- such as research consortia,
public-private partnerships, or organisational alliances -- would help assess
whether the observed patterns of architectural brokerage and connectedness
represent broader organisational regularities. These extensions would further
integrate the framework into the comparative study of boundary-spanning
architectures in complex sociotechnical systems.
 
\backmatter

\bmhead{Acknowledgements}

Acknowledgements are not compulsory. Where included they should be brief. Grant
or contribution numbers may be acknowledged.

Please refer to Journal-level guidance for any specific requirements.

\section*{Declarations}

Some journals require declarations to be submitted in a standardised format.
Please check the Instructions for Authors of the journal to which you are
submitting to see if you need to complete this section. If yes, your manuscript
must contain the following sections under the heading `Declarations':

\begin{itemize}
\item Funding
\item Conflict of interest/Competing interests (check journal-specific
guidelines for which heading to use)
\item Ethics approval and consent to participate
\item Consent for publication
\item Data availability 
\item Materials availability
\item Code availability 
\item Author contribution
\end{itemize}

\bibliography{references.bib}

\end{document}